\documentstyle[12pt]{article}

\newcommand{\be}{\begin{equation}}
\newcommand{\ee}{\end{equation}}
\newcommand{\bea}{\begin{eqnarray*}}
\newcommand{\eea}{\end{eqnarray*}}
\newcommand{\ba}{\begin{eqnarray}}
\newcommand{\ea}{\end{eqnarray}}
\newcommand{\bee}{\begin{enumerate}}
\newcommand{\ene}{\end{enumerate}}

\begin{document}

{\Large \bf Remarkable Mass Relations in the Electroweak Model
}

\vskip 2cm

\begin{center}
{\large Jean Pestieau}\\
Institut de Physique Th\'eorique, Universit\'e catholique de
Louvain,\\ Chemin du Cyclotron 2, B-1348 Louvain-la-Neuve,
Belgium\footnote{pestieau@fyma.ucl.ac.be}
\end{center}

\vskip 2cm

\begin{abstract}
In the electroweak standard model we observe two remarkable
empirical mass relations, $m_W + m_B = {v\over 2}$ and $m_W - m_B
= e{v\over 2}$ where $m^2_Z \equiv m^2_W + m^2_B$, $e$ is the
positron electric charge and $v$, the strength of the Higgs
condensate.
\end{abstract}

\newpage

\par\noindent
The electroweak standard model \cite{1, 2, 3} is
based on the gauge group $SU(2) \otimes U(1)$, with
gauge bosons fields $W^{i}_\mu = 1, 2, 3$ and
$B_\mu$. After spontaneous symmetry breaking,
$W^{i}_\mu$ and $B_\mu$ acquire masses, $m_W$ and
$m_B$ respectively. $W^3_\mu$ and $B_\mu$ are mixed.
After diagonalization of mass matrix, one gets the
physical fields $A_\mu$ and $Z_\mu$ corresponding to
the massless photon and the neutral $Z$ boson of
mass $m_Z$, with  the relations \cite{2, 4}

\ba
m^2_Z &=& m^2_W + m^2_B \label{1}\\
\cos \theta_W &=& {m_W \over m_Z} \ \ \ , \ \ \ \sin
\theta_W = {m_B \over m_Z}\label{2}
\ea
where $\theta_W$ is the weak mixing angle :
\ba
Z_\mu &=& W^3_\mu \cos \theta_W - B_\mu \sin
\theta_W \label{3}\\
A_\mu &=& W^3_\mu \sin \theta_W + B_\mu \cos
\theta_W \label{4}
\ea
>From experimental data, we get two remarkable
empirical mass relations defining two different mass
scales :
\ba
m_W + m_B &=& {v\over 2} \label{5}\\
m_W - m_B &=& e {v\over 2} \label{6}
\ea
where $e$ is the positron electric charge and $v$, the
strength of the Higgs condensate.
\\
Eqs (\ref{2}), (\ref{5}) and (\ref{6}) imply \cite{5} :
\be
e = {1-\mbox{tg} \  \theta_W \over 1 + \ \mbox{tg} \
\theta_W} = \mbox{tg} \ \left({\pi\over 4} - \theta_W \right)
\label{7}
\ee
>From the experimental data \cite{6}
\ba
&&\alpha = {e^2\over 4\pi} = {1\over
137.03599976(50)} \label{8}\\
&&m_Z = 91.1872 \pm 0.0021 \ \mbox{GeV}, \label{9}
\ea
we obtain :
\ba
\mbox{tg} \ \theta_W &=& 0.5351 \label{10}\\
m_W &=& {v\over 4} (1+e) =  80.400 \pm 0.002 \ \mbox{GeV}
\label{11}\\
m_B &=& {v\over 4} (1-e) = 43.022 \pm 0.001 \ \mbox{GeV}
\label{12}\\ v &=& 246.85 \ \mbox{GeV} \label{13}
\ea
to be compared with the experimental values \cite{6} :
\be
m_W (\exp) = 80.419 \pm 0.038 \ \mbox{GeV}
\label{14}
\ee
\be
v_F \equiv \left({1\over \sqrt{2} G_F}\right)^{1/2}
= 246.22 \ \mbox{GeV} \label{15}
\ee
$v_F$ is the usual value of the strength of the
Higgs condensate obtained from $G_F$, the Fermi
constant derived from the muon lifetime \cite{6} and
\be
{v\over v_F} - 1 = 0.0026 \label{16}
\ee
We conclude that Eqs (\ref{5}) and (\ref{6}) are
robust relations and point to new physics.

\par\noindent
With respect to weak $SU(2), W^{i}_\mu$'s
transform as a weak isotriplet, $B_\mu$ as a weak
singlet. Let us define
\ba
D_\mu &=& {1\over \sqrt{2}} (W^3_\mu -
B_\mu)\label{17}\\
U_\mu &=& {1\over \sqrt{2}} (W^3_\mu +
B_\mu).\label{18}
\ea
$W^+_\mu, W^-_\mu, D_\mu$ and $U_\mu$ are the fields
obtained from the direct product of two isodoublet
fundamental fields. Then the masses of $D$ and $U$
fields are
\ba
m_D &=& {1\over \sqrt{2}} (m_W + m_B) = {v\over
2\sqrt{2}} = 87.27 \ \mbox{GeV} \label{19}\\
m_U &=& {1\over \sqrt{2}} (m_W - m_B) = {ev\over
2\sqrt{2}} = 26.43 \ \mbox{GeV}. \label{20}
\ea
Now we can see the pattern of the neutral vector
boson mass matrix :
\ba
2 {\cal L}_M &=& \mbox{\Huge{$
\makebox[0mm]{\raisebox{3mm}[0mm][0mm]{\hspace*{14mm}$\frown$}}$
{\small$Z_\mu A_\mu$}
\makebox[0mm]{\raisebox{-5mm}[0mm][0mm]{\hspace*{-15mm}$\smile
$}}$
$}} \left(\begin{array}{cccc} m^2_Z & 0 \\ 0 & 0 \end{array}\right)
\left(\begin{array}{c} Z_\mu \\ A_\mu \end{array}\right) = \nonumber
\\
&&\mbox{\Huge{$
\makebox[0mm]{\raisebox{3mm}[0mm][0mm]{\hspace*{14mm}$\frown$}}$
{\small$W^3_\mu B_\mu$}
\makebox[0mm]{\raisebox{-5mm}[0mm][0mm]{\hspace*{-17mm}$\smile
$}}$
$ }}
\left(\begin{array}{lll} m^2_W & -m_W m_B \\ -m_W m_B &
m^2_B\end{array}\right) \left(\begin{array}{c}W^3_\mu \\ B_\mu
\end{array}\right) =
\nonumber \\
&& \mbox{\Huge{$
\makebox[0mm]{\raisebox{3mm}[0mm][0mm]{\hspace*{14mm}$\frown$}}$
{\small$D_\mu U_\mu$}
\makebox[0mm]{\raisebox{-5mm}[0mm][0mm]{\hspace*{-16mm}$\smile
$}}$
$ }}
\left(\begin{array}{ccc} m^2_D & m_Dm_U \\ m_Dm_U & m^2_U
\end{array}\right) \left(\begin{array}{c} D_\mu \\ U_\mu
\end{array}\right)
\nonumber\\
&&\mbox{\Huge{$
\makebox[0mm]{\raisebox{3mm}[0mm][0mm]{\hspace*{14mm}$\frown$}}$
{\small$D_\mu U_\mu$}
\makebox[0mm]{\raisebox{-5mm}[0mm][0mm]{\hspace*{-16mm}$\smile
$}}$
$ }}
{v^2\over 8} \left(\begin{array}{ccc} 1 & e \\ e & e^2
\end{array}\right) \left(\begin{array}{c} D_\mu \\ U_\mu
\end{array}\right)
\label{21}
\ea
It is an empirical fact that the electric charge
sets the mass scale between $U_\mu$ and $D_\mu$ fields:
\be
e = {m_U \over m_D} = \mbox{tg} \ \left({\pi\over 4} -
\theta_W\right).
\label{22}
\ee

\par\noindent
It is interesting to note that
\be
m_t = 174.3 \pm 5.1 \ \mbox{GeV} \label{23}
\ee
to be compared with the empirical relation
\be
m_t = {v\over \sqrt{2}} = \sqrt{2} (m_W + m_B) = 174.6 \
\mbox{GeV} \label{24}
\ee
and comparing Eqs (\ref{24}) and (\ref{19}), we find
\be
{m_t \over m_D} = 2
\label{25}
\ee

\end{document}